\documentclass{mem}
\usepackage{natbib}
\usepackage{txfonts}
\usepackage{balance}
\usepackage{graphicx}
\usepackage[a4paper,breaklinks,dvipdfm]{hyperref}
\idline{00}{000}
\begin{document}
\def\ergps{erg~s$^{-1}$}
\def\teff{$T\rm_{eff }$}
\def\kms{$\mathrm {km s}^{-1}$}
\def\Msun{\,M_\odot}
\def\arcsec{^{\prime\prime}}
\def\ltsim{\raise 2pt \hbox {$<$} \kern-1.1em \lower 4pt \hbox {$\sim$}}
\def\ltapprox{\raise 2pt \hbox {$<$} \kern-1.1em \lower 5pt \hbox {$\approx$}}
\def\gtsim{\raise 2pt \hbox {$>$} \kern-1.1em \lower 4pt \hbox {$\sim$}}
\def\gtapprox{\raise 2pt \hbox {$>$} \kern-1.1em \lower 5pt \hbox {$\approx$}}
\def\arcsec{$^{\prime\prime}$}
\def\arcmin{$^{\prime}$}
\def\degrees{$^{\circ}$}
\def\etal{{\it et al.~}}
\def\Ang{$A^{\circ}$}
\def\phat{\hat{p}}
\def\zw{ZwCl~2341.1+0000\,}
\title{
A deep radio and X-ray view of cluster formation at the crossroads of filaments
}
   \subtitle{}

\author{
J. \,Bagchi\inst{1} 
\and R. J. van Weeren\inst{2}
\and S. Raychaudhury\inst{3}
\and H. J. A. R\"ottgering\inst{2}
\and H. T. Intema\inst{2}
\and F. Miniati\inst{4}
\and T. A. En\ss{}lin\inst{5}
\and M. Markevitch\inst{6}
\and{T. Erben}\inst{7}
          }


\institute{
Inter-University Centre for Astronomy and Astrophysics (IUCAA), 
Pune 411007, India
\and
Leiden Observatory, Leiden University, PO Box 9513, 2300 RA Leiden, The Netherlands
\and
School of Physics and Astronomy, University of Birmingham, Edgbaston, Birmingham B15 2TT, UK 
\and
Physics Department, Wolfgang-Pauli-Strasse 27,  ETH-Z{\"u}rich, CH-8093 Z{\"u}rich, Switzerland
\and
Max-Planck-Institut f{\"u}r Astrophysik, Karl-Schwarzschild-Str.1, PO Box 1317, 85741 Garching, Germany
\and
Harvard-Smithsonian Center for Astrophysics, 60 Garden Street, Cambridge, MA 02138, USA
\and
Argelander-Institut f{\"u}r Astronomie, Universit{\"a}t Bonn, Auf dem H{\"u}gel 71, D-53121 Bonn, Germany
\vskip 0.1cm
\email{joydeep@iucaa.ernet.in}
}

\authorrunning{Bagchi et al.}

\titlerunning{ZwCl~2341.1+0000: deep radio and X-ray view}

\abstract{
Deep X-ray data from {\it Chandra} and {\it XMM-Newton}, and GMRT radio data are 
presented for \zw, an extremely unusual and  complex merging cluster of galaxies at the
intersection of optical filaments. We propose that energetics of multiple mergers and accretion flows 
has resulted in wide-spread shocks, acceleration of cosmic ray particles and 
amplification of weak magnetic fields. This results in Mpc-scale peripheral radio relics 
and halo like non-thermal emission observed near the merging center.

}
\maketitle{}

\section{Introduction}
Galaxy clusters grow by mergers of smaller subclusters and galaxy groups as predicted by
hierarchical models of large-scale structure (LSS) formation. For the most massive mergers a significant
amount of energy is released; upto $10^{63} - 10^{64}$ erg, according to these models. 
All massive clusters have undergone several mergers in their history and
presently clusters are still growing by matter accretion at the junction of large
scale filaments of galaxies, mediated by the gravity of dark-matter component. Key properties for testing
models of LSS formation include the total energy budget and the
detailed temperature distribution within a cluster, which are both strongly affected by the
cluster's merger history. Moreover, the physics of shock waves in the tenuous intra-cluster
medium (ICM) and the effect of cosmic-rays (CR; relativistic particles) on galaxy clusters
are all fundamental for our understanding of LSS formation.

Diffuse radio sources with steep spectra ($\alpha \lesssim -0.5$), not
directly associated with individual galaxies, are observed in a number of clusters ( cf. review
by \citet{Ferrari_08}). Large scale ($l \gtrsim500$ kpc) diffuse radio sources in clusters are commonly
divided into `radio halos'  and `radio relics'. Radio halos have smooth morphologies, are extended
with sizes $\gtrsim$1 Mpc, unpolarized, and are found at the centers of clusters, co-spatial with the
X-ray emitting hot ICM gas. Giant radio relics are found on  the cluster periphery,
with sizes up to several Mpc, sometimes showing arc or ring-like structures, and are
highly polarized (p$\sim$ 10-$50 \%$ at 1.4 GHz). They are probably signatures of electrons
accelerated at large-scale shocks. The vast majority of clusters with extended diffuse radio sources
are massive, X-ray luminous and are merging systems. Although still under debate, shocks
and turbulence, both caused by  mergers, are thought to be responsible for (re)accelerating electrons to
highly relativistic energies, and synchrotron radiation is produced in the presence of 
magnetic fields in the ICM. Therefore deep radio and X-ray observations of merging clusters 
may provide extremely important information on the dynamics of LSS formation.

\begin{figure*}[t!]
\resizebox{\hsize}{!}{\includegraphics[clip=true]{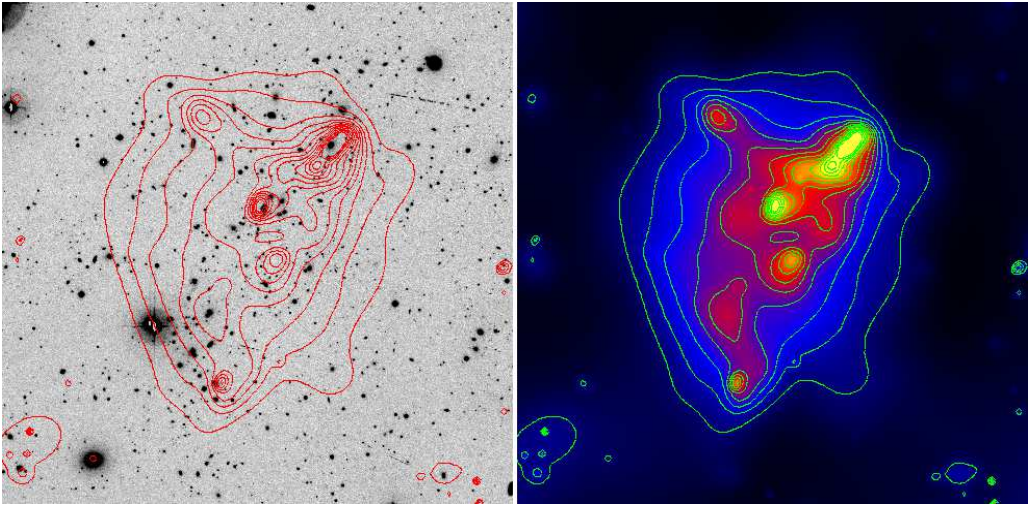}}
\caption{\footnotesize
Left panel: The optical R-band image of
merging system \zw obtained from WFI at the
ESO 2.2m telescope. The overlaid contours show the  thermal X-ray
emission in the energy range 0.5--3~keV observed with {\it XMM-Newton}.
Right panel: Grey-scale version of the same X-ray emission  observed with {\it XMM-Newton}.
Both images cover about 4~Mpc on a side at redshift z=0.27 
(\citet{Raychaudhury_et_al_2011}).
}
\label{xmm}
\end{figure*}
\zw, discovered by \citet{Bagchi_et_al_2002}, is
an extremely interesting and  complex system of galaxies,
composed out of several different subclusters arranged along a north-south filamentary axis extending
upto at least $\sim$4 Mpc. It is probably a very large and massive cluster 
coming together at the junction of supercluster filaments. A 2.5 Mpc scale diffuse radio emission 
was detected in the main filamentary structure \citep{Bagchi_et_al_2002,Giovannini_et_al_2010} 
which makes it unique and a highly valuable source for cosmological studies.
The system is located at a redshift of $z=0.27$, based on SDSS DR7 spectroscopic redshifts 
of several galaxies in the vicinity. A galaxy isodensity map
derived from SDSS imaging data shows an elongated cluster of galaxies, including several
subclusters distributed roughly along a north-south axis, spanning $\gtrsim$4~Mpc.
A network of  several filaments of galaxies are seen branching off from the 
main structure towards the east and northeast forming a `cosmic-web' (Fig.~\ref{xmm} \& Fig.~\ref{gmrt}).

\section{Radio and X-ray Observations of \zw}

\begin{figure}[]
\resizebox{\hsize}{!}{\includegraphics[clip=true]{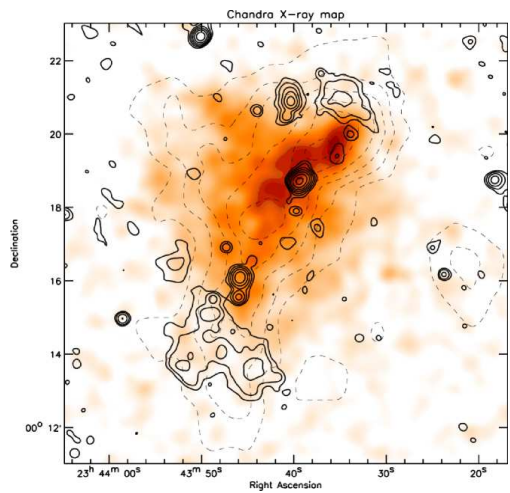}}
\caption{
\footnotesize
A radio, optical and X-ray overlay map. The solid contours represent the radio
emission at 610 MHz from the GMRT data [\citet{van_Weeren_2009}]. Broad beam
of 15 $^{\prime \prime }$ FWHM is used to show better the diffuse radio emission. 
The radio contours are drawn
at  $[1, 2, 4, 8, 16, 32, ...] \times 0.224$ mJy/beam. Dashed contours show the galaxy
isodensity contours from SDSS, corresponding roughly to a limit of  $M_{{\rm V}} = -18.1$
(i.e., $M^\ast + 2.4$). The dashed contours are drawn at [2, 3, 4, 5,...] galaxies arcmin$^{-2}$ with
a redshift cut.  Grey-scle image is smoothed, point source subtracted X-ray emission map from {\em Chandra}
in the 0.5-3.0 keV energy band.
}
\label{gmrt}
\end{figure}

For probing the Bremsstrahlung emission from the diffuse, hot ICM of this remarkable large-scale structure,
deep X-ray observations with {\it Chandra} and {\it XMM-Newton} were obtained. The X-ray emission 
is detected for about 2.5 Mpc in the N - S direction and an extension
towards the east is also visible, similar to the pattern of galaxy distribution. 
In Fig.~\ref{gmrt} we also show the GMRT 610~MHz radio
contours superposed on the {\it Chandra} ACIS-I X-ray image. In addition to N - S X-ray emission 
roughly following the filament of galaxies visible in the optical image, embedded
within the extended diffuse emission, we can identify several other
prominent X-ray emitting systems, which  most likely represent hot gas associated with
merging groups  and clusters (Fig.~\ref{xmm}). 

The overall cluster temperature in a $r\!=\! 4^\prime$ region
(roughly 1~Mpc or  close to $r_{500}$), measured from the 
Chandra ACIS-I observation, is $4.4\pm 0.6$~keV.  The corresponding X-ray luminosity within the
same region in the 0.1--2.4~keV range is $L_X\!=\! 2.4\times 10^{44}$ \ergps, and
the bolometric X-ray luminosity being $L_{bol}\!=\! 4.6\times 10^{44}$ \ergps ($h=0.71$,
$\Omega_{M}=0.27$, $\Omega_{\Lambda}=0.73$). From the XMM-Newton combined PN and MOS observations, 
from the central $r\!=\! 4^\prime$ region, we obtain an average temperature of $4.7\pm 0.5$~keV. 
Thus both Chandra and XMM-Newton measured temperatures are consistent with each other. 

Deep GMRT radio maps  were published by us recently \citep{van_Weeren_2009} at 610, 240 and 150 MHz. 
These maps (specially 610 MHz map shown here in Fig.~\ref{gmrt})
successfully resolved the compact sources from the extended
diffuse emission of extremely low surface brightness (only 100 - 200 $\mu$Jy/beam at 610 MHz),
which was found to be located mainly in two Mpc-scale
radio structures to the north and south of the cluster, capping the two ends of X-ray emitting hot gas  and the
central optical filament. We interpret this structure as being a peripheral double radio relic, 
where the particles are accelerated by the diffusive shock acceleration (DSA or Fermi-I)  mechanism in outward
propagating shock waves generated in  major cluster merger activity observed at the center (cf. \citet{Bykov_08}). 
The merger axis is clearly along the galaxy filament, with radio relics placed tangential 
to this axis, as expected in a merger-shock geometry.
However, to the limits of surface-brightness sensitivity achieved ($\sim30$~$\mu$Jy noise 
for a $5^{\prime\prime}$ beam at 610 MHz, $\sim$12h on-source time),
no central radio-halo was detected in our  GMRT observations. Possible reason is that some of the 
short baselines in the GMRT observations were affected by RFI and had to be flagged,
which allowed imaging of spatial scales of $< 4^{\prime}$ only (at 610 MHz). A very deep 327 MHz 
GMRT observation is in preparation  to resolve this issue.

\section{Discussion}
The X-ray images clearly reveal a highly unrelaxed cluster
experiencing a complex merger that is occurring
preferentially along the NW-SE axis -- i.e., along the major
filament clearly seen in the distribution of galaxies. An additional arm of the 
X-ray emission extends from the middle of the cluster towards NE, roughly coinciding 
with another filamentary galaxy structure. ZwCl~2341.1+0000 may thus represent by 
far the most complex merger configuration found among
galaxy clusters. Due to its highly unrelaxed dynamical state revealed by our
X-ray data, this system may provide a rare glimpse into the early steps of
assembly of a galaxy cluster at the junction of filaments. In addition, rare detection
of diffuse radio emission from such a  system will throw much light on the dynamics of
cluster formation.  The optical, radio and X-ray
observations presented in this paper firmly establishes the link of
this merger to the axis of a supercluster filament along which the
merger takes place. This work represents a substantial advance in the field,
because it probes several important components of the cosmic environment: intergalactic gas,
magnetic fields, and cosmic-rays.  Large scale radio emission indicates that magnetic fields of appreciable
strength are present not only in the ICM but also in the diffuse intergalactic medium, i.e., in the gas
that will be shocked as it accretes onto collapsing structures - the precursors of virialized galaxy clusters.
The search for magnetic fields in the intergalactic medium is of fundamental importance for cosmology
\citep{Bonafede_et_al_ 2010,Dolag_2006}.

In a recent deep 1.4~GHz VLA observation, \cite{Giovannini_et_al_2010} present evidence
of 2 Mpc-scale polarized radio emission, midway between the peripheral radio relics. 
The radio emission clearly follows the distribution
of optical galaxies and X-ray emission which might
suggest that it is similar to  radio halos found in dynamically unrelaxed clusters \citep{Ferrari_08}.
The polarization vectors (p$\sim$10 - 20\%) are very regular, 
indicating an organized magnetic field pattern.
This feature might result from the accretion flow shocks in filaments
compressing the weak magnetic fields of the infalling IGM, which is consistent with  a model proposed by
Bagchi et al. (2002). An observation of large scale diffuse radio emission  and magnetic fields 
in filamentary galactic environment  also  provides a foretaste of future detections, as a major goal 
of upcoming  low frequency radio telescopes like the LOFAR and LWA is
to probe the non-thermal processes in hundreds of  merging clusters and
in the filamentary cosmic-web of the near and distant Universe.

\section{Conclusions}

We have presented deep X-ray and radio observations which show that \zw\, constitutes a 
complex,  large-scale ensemble of clusters, 
which  are merging  at the intersection of filaments identifiable in optical
and X-ray images. Energetics of multiple mergers and accretion flows has resulted 
in wide-spread shocks, acceleration of cosmic ray particles and amplification of weak magnetic fields.
This is evident in a pair of giant, peripheral radio relics  and a faint, 
diffuse radio halo extending more than $\sim$1 Mpc between the two relics, following the
large-scale filamentary structure.  

\begin{acknowledgements}
We thank the operations team of the NCRA--TIFR GMRT observatory for help which resulted in
the radio data. The GMRT is a national facility operated by the National Centre for 
Radio Astrophysics of the TIFR, India. 
\end{acknowledgements}

\bibliographystyle{aa}

\end{document}